\documentclass[conference,10pt]{IEEEtran}
\IEEEoverridecommandlockouts
\renewcommand{\baselinestretch}{0.99}

\usepackage{algorithm,algorithmic}
\makeatletter
\newcommand\fs@betterruled{%
  \def\@fs@cfont{\bfseries}\let\@fs@capt\floatc@ruled
  \def\@fs@pre{\vspace*{5pt}\hrule height.8pt depth0pt \kern2pt}%
  \def\@fs@post{\kern2pt\hrule\relax}%
  \def\@fs@mid{\kern2pt\hrule\kern2pt}%
  \let\@fs@iftopcapt\iftrue}
\floatstyle{betterruled}
\restylefloat{algorithm}

\makeatother

\usepackage{cite}
\usepackage{amsmath,amssymb,amsfonts}
\usepackage{algorithm,algorithmic}
\usepackage{graphicx}
\usepackage{amsmath}
\usepackage{textcomp}
\usepackage{comment}
\usepackage{enumitem, subcaption}
\usepackage{xcolor}
\usepackage{color,soul}
\def\BibTeX{{\rm B\kern-.05em{\sc i\kern-.025em b}\kern-.08em
    T\kern-.1667em\lower.7ex\hbox{E}\kern-.125emX}}
\renewcommand{\a}{\mathbf{a}}

\newcommand{\e}{\mathbf{e}}

\newcommand{\h}{\mathbf{h}}

\newcommand{\n}{\mathbf{n}}

\newcommand{\p}{\mathbf{p}}

\newcommand{\s}{\mathbf{s}}

\renewcommand{\v}{\mathbf{v}}
\newcommand{\w}{\mathbf{w}}
\newcommand{\x}{\mathbf{x}}
\newcommand{\y}{\mathbf{y}}

\newcommand{\D}{\mathbf{D}}

\newcommand{\F}{\mathbf{F}}

\renewcommand{\H}{\mathbf{H}}
\newcommand{\I}{\mathbf{I}}

\newcommand{\K}{\mathbf{K}}

\newcommand{\N}{\mathbf{N}}

\renewcommand{\P}{\mathbf{P}}

\newcommand{\R}{\mathbf{R}}
\renewcommand{\S}{\mathbf{S}}

\newcommand{\V}{\mathbf{V}}
\newcommand{\W}{\mathbf{W}}

\newcommand{\Y}{\mathbf{Y}}

\newcommand{\Compl}{\mbox{$\mathbb{C}$}}

\renewcommand{\Re}{\mathrm{Re}}

\usepackage{xcolor, soul}
\sethlcolor{yellow}
\DeclareMathAlphabet\mathbfcal{OMS}{cmsy}{b}{n}

\begin{document}
\title{Simultaneous Near-Field THz Communications and Sensing with Full Duplex Metasurface Transceivers}
\author{Ioannis Gavras$^{1}$ and George C. Alexandropoulos$^{1,2}$\\
$^1$Department of Informatics and Telecommunications, National and Kapodistrian University of Athens, Greece
\\$^2$Department of Electrical and Computer Engineering, University of Illinois Chicago, IL, USA
\\e-mails: \{giannisgav, alexandg\}@di.uoa.gr
\thanks{This work has been supported by the SNS JU projects TERRAMETA and 6G-DISAC under the EU's Horizon Europe research and innovation program under Grant Agreement numbers 101097101 and 101139130, respectively.} \vspace{-0.45cm}
}

\maketitle
\begin{abstract}
In this paper, a Full Duplex (FD) eXtremely Large (XL) Multiple-Input Multiple-Output (MIMO) node equipped with reconfigurable metasurface antennas at its transmission and reception sides is considered, which is optimized for simultaneous multi-user communications and sensing in the near-field regime at THz frequencies. We first present a novel Position Error Bound (PEB) analysis for the spatial parameters of multiple targets in the vicinity of the FD node, via the received backscattered data signals, and devise an optimization framework for its metasurface-based precoder and combiner. Then, we formulate and solve an optimization problem aiming at the downlink sum-rate maximization, while simultaneously ensuring a minimum PEB requirement for targets' localization. Our simulation results for a sub-THz system setup validate the joint near-field communications and sensing capability of the proposed FD XL MIMO scheme with metasurfaces antennas, showcasing the interplay of its various design parameters.
\end{abstract}

\begin{IEEEkeywords}
Metasurfaces antennas, XL MIMO, full duplex, integrated sensing and communications, near field, THz.
\end{IEEEkeywords}

\section{Introduction}
The upcoming sixth Generation (6G) of wireless networks is expected to offer advanced radar-style sensing capabilities in an manner integrated with the evolution of 5G's use cases\cite{6G-DISAC_mag}. This Integrated Sensing and Communications (ISAC) paradigm~\cite{mishra2019toward} is lately receiving remarkable research and standardization attention, with emphasis being given on use cases, channel modeling, and enabling technologies. To this end, the combination of eXtremely Large (XL) Multiple-Input Multiple-Output (MIMO) systems with THz frequencies promises high angular and range resolution~\cite{THs_loc_survey}, reconfigurable intelligent surfaces enable programmable correlations of sensing and communication subspaces~\cite{RIS_ISAC}, and in-band Full Duplex (FD) radios allow for simultaneous communications and monostatic sensing via data signal transmissions~\cite{FD_JSAC_Survey}. 

The research in FD-enabled ISAC systems ranges from single-antenna setups \cite{barneto2019full,liyanaarachchi2021optimized} to massive MIMO configurations at millimeter-wave frequencies \cite{barneto2020beamforming,Islam_2022_ISAC,Atiq_ISAC_2022,FD_MIMO_VTM2022,bayraktar2023hybrid}, considering single or multiple radar targets and User Equipment (UE). Very recently, \cite{FD_HMIMO_2023,gavras2024joint} presented an FD XL MIMO architecture with Transmit/Receive (TX/RX) Dynamic Metasurface Antennas (DMAs)~\cite{Shlezinger2021Dynamic} that was designed for simultaneous near-field THz communications and sensing. To the best of the authors' knowledge, all current ISAC formulations consider communication metrics in conjunction with radar gain indicators, either as dual optimization objectives or in interchangable roles between a core optimization objective and a constraint.

In this paper, focusing on the FD-enabled XL MIMO ISAC system model of~\cite{FD_HMIMO_2023}, we consider a DMA-based FD node operating at THz frequencies wishing to estimate the $3$-Dimensional (3D) position of multiple targets in its near-field vicinity via the backscattered data signals transmitted in the DownLink (DL) direction to a subset of them operating as network-connected UEs. We propose a novel ISAC problem formulation targeting the DL sum-rate maximization under a maximum threshold for the targets' Position Error Bound (PEB), describing the accuracy of their 3D spatial parameters' estimation. A PEB analysis tailored to targets' signal reflections received via the considered FD node's DMA-based RX is presented. To tackle the proposed ISAC objective, we devise a scheme that seeks to identify the best combination between the sensing- and communication-optimized system parameters. Selective numerical evaluations for a sub-THz system setup are provided showcasing the validity of the conducted PEB analysis as well as the combined superior PEB and DL rate performance compared to state-of-the-art strategies. 

\textit{Notations:} Vectors and matrices are denoted by boldface lowercase and boldface capital letters, respectively. The transpose, Hermitian transpose, and the inverse of $\mathbf{A}$ are denoted by $\mathbf{A}^{\rm T}$, $\mathbf{A}^{\rm H}$, and $\mathbf{A}^{-1}$, respectively, while $\mathbf{I}_{n}$ and $\mathbf{0}_{n}$ ($n\geq2$) are the $n\times n$ identity and zeros' matrices, respectively. $[\mathbf{A}]_{i,j}$ is the $(i,j)$-th element of $\mathbf{A}$, $\|\mathbf{A}\|$ returns $\mathbf{A}$'s Euclidean norm, $|a|$ is the amplitude of a complex scalar $a$, $\mathbb{C}$ is the complex number set, and $\jmath$ is the imaginary unit. $\mathbb{E}\{\cdot\}$ is the expectation operator and $\mathbf{x}\sim\mathcal{CN}(\mathbf{a},\mathbf{A})$ indicates a complex Gaussian random vector with mean $\mathbf{a}$ and covariance matrix $\mathbf{A}$.

\section{System and Signal Models}\label{sec: system_signal}
\subsection{ISAC System Model}
Consider an FD-enabled ISAC system realized via an FD XL MIMO node as in \cite[Fig. 1]{FD_HMIMO_2023}  missioned to localize $K$ targets in its near-field region, $U\leq K$ out of which are also served with data in DL direction. The FD node is assumed equipped with TX and RX DMAs~\cite{Shlezinger2021Dynamic} and the $U$ targets are single-antenna communication UEs. Each of TX/RX DMAs consists of $N_{\rm RF}$ single-RF-fed microstrips, where transmitted/received signals are phase-controlled via $N_{\rm E}$ metamaterials with dynamically tunable frequency responses, thus implementing analog BeamForming (BF). We define $N\triangleq N_{\rm RF}N_{\rm E}$ and use notation $d_{\rm E}$ to denote the inter-metamaterial spacing in TX/RX DMAs.

The DMA-based TX possesses the unit-powered complex-valued symbol $\s_{u}$ for each $u$th UE which is first digitally precoded via $\v_u\in\Compl^{N_{\rm RF}\times 1}$. Before transmission in the DL, all $U$ precoded symbols are analog processed via the weights of the TX DMA, resulting in the $N$-element transmitted signal $\x\triangleq\P_{\rm TX}\W_{\rm TX}\V\s$, where $\V \triangleq [\v_1,\ldots,\v_U]\in\Compl^{N_{\rm RF}\times U}$, $\s \triangleq [s_1,\ldots,s_U]^{\rm T}\in\Compl^{U\times1}$, $\P_{\rm TX}$ denotes the $N\times N$ matrix modeling the signal propagation inside the microstrips at the TX~\cite[eq.~(1)]{Xu_DMA_2022} (we define similarly the $N\times N$ matrix $\P_{\rm RX}$ for the FD node's DMA-based RX), and $\W_{\rm TX}\in\mathbb{C}^{N\times N_{\rm RF}}$ represents the analog TX BF matrix, which is defined $\forall$$i=1,\dots,N_{\rm RF}$ and $\forall$$n = 1,\dots,N_{\rm E}$ as follows~\cite{FD_HMIMO_2023}:
\begin{align}
    [\W_{\rm TX}]_{(i-1)N_{\rm E}+n,j} = \begin{cases}
    w^{\rm TX}_{i,n}\in \mathcal{W},&  i=j\\
    0,              & i\neq j
\end{cases}
\end{align}
with $\mathcal{W}\triangleq \left\{0.5(\jmath+e^{\jmath\phi})|\phi\in\left[-0.5\pi,0.5\pi\right]\right\}$. Similarly, we define the weights $w^{\rm RX}_{i,n} \in \mathcal{W}$ $\forall$$i,n$, from which the analog RX combiner  $\W_{\rm RX}\in\mathbb{C}^{N\times N_{\rm RF}}$ is formulated. We finally assume that the transmitted data signal, which is also used for monostatic sensing at the FD XL MIMO node, is power limited such that $\mathbb{E}\{\|\P_{\rm TX}\W_{\rm TX}\V\s\|^2\}\leq P_{\rm max}$, where $P_{\rm max}$ denotes the maximum transmission power.

\subsection{Near-Field Channel Model}
Each $1\times N$ complex-valued DL THz channel (i.e., for each $u$th UE) is modeled as follows:
\begin{align}
    \label{eqn:DL_chan}
    [\h_{{\rm DL},u}]_{(i-1)N_{\rm E}+n} \triangleq \alpha_{u,i,n} \exp\left(\frac{\jmath2\pi}{\lambda} r_{u,i,n}\right),
\end{align}
where $\alpha_{u,i,n}$ models the attenuation factor including molecular absorption~\cite[eq.~(5)]{FD_HMIMO_2023}, $r_{u,i,n}$ represents the distance between the $u$th UE's antenna and the $n$th meta-element of each $i$th TX DMA's Radio Frequency (RF) chain, and $\lambda$ is the wavelength. 

We consider $K$ targets with spherical coordinates $\{(r_1,\theta_1,\varphi_1),\ldots,(r_K,\theta_K,\varphi_K)\}$, including the distances from the origin, and the elevation and azimuth angles, respectively. From those targets, $U$ out of $K$ with coordinates $\{(r_1,\theta_1,\varphi_1),\dots,(r_U,\theta_U,\varphi_U)\}$ are the DL UEs. Each distance $r_{u,i,n}$ in \eqref{eqn:DL_chan} can be calculated as:
\begin{align}\label{eq: dist}
    \nonumber &r_{u,i,n}\! =\! \!\Big(\!(r_{u}\sin\theta_{u}\cos\varphi_{u} +\frac{d_{\rm P}}{2}+(i\!-\!1)d_{\rm RF})^2 \\ &+(r_{u}\sin\theta_{u}\sin\varphi_{u})^2 + (r_{u}\cos\theta_{u}\!-\!(n\!-\!1)d_{\rm E})^2\Big)^{\frac{1}{2}}
\end{align}
with $d_{\rm P}$ being the distance between TX/RX DMAs and $d_{\rm RF}$ is the distance between any two adjacent TX or RX microstrips.

The end-to-end MIMO channel between the RX/TX DMAs of the FD XL MIMO node via the reflections from all $K$ targets in the node's vicinity, when all those entities are considered as point sources with coordinates $(r_k,\theta_k,\varphi_k)$ $\forall$$k=1,2,\ldots,K$, can be expressed as follows: 
\begin{align}\label{eq:H_R}
    \H_{\rm R} \triangleq \sum\limits_{k=1}^{K}\beta_k \a_{\rm RX}(r_k,\theta_k,\varphi_k)\a_{\rm TX}^{\rm H}(r_k,\theta_k,\varphi_k)
\end{align}
where $\beta_k$ is the complex-valued reflection coefficient for each $k$th target and $\a_{\rm str}(\cdot)$, with ${\rm str}\triangleq\{{\rm TX},{\rm RX}\}$, is defined as:
\begin{align}
    \label{eq:response_vec}
    [\a_{\rm str}(r_k,\theta_k,\phi_k)]_{(i-1)N_E+n} \triangleq \alpha_{k,i,n}\exp\Big({\jmath\frac{2\pi}{\lambda}r_{k,i,n}}\Big),
\end{align}
where $r_{k,i,n}$ is computed analogously to \eqref{eq: dist}, adjusting for either the TX-UE or the UE-RX distance, as appropriate.


\subsection{Received Signal Models}
The baseband received signal $y_{u}\in\Compl$ at each $u$th UE at each $t$th time slot can be mathematically expressed as:
\begin{align}
    y_{u}(t) \triangleq \h_{{\rm DL},u}\P_{\rm TX}\W_{\rm TX}\v_u s_u(t) + n_u(t),
\end{align}
where $n_u(t)\sim\mathcal{CN}(0,\sigma^2)$ denotes the Additive White Gaussian Noise (AWGN) vector. Assuming we employ $T$ transmissions for each communication slot, where $t\in\{1,\ldots, T\}$, the mathematical expression for the baseband received signal $\Y\in\Compl^{N_{\rm RF}\times T}$ at the output of the RX DMA, is as follows:
\begin{align}\label{eq:received_matrix}
    \Y \triangleq& \W^{\rm H}_{\rm RX}\P^{\rm H}_{\rm RX}\H_{\rm R}\P_{\rm TX}\W_{\rm TX}\V\S \\
    &\nonumber+ (\W^{\rm H}_{\rm RX}\P^{\rm H}_{\rm RX}\H_{\rm SI}\P_{\rm TX}\W_{\rm TX} + \D)\V\S + \W^{\rm H}_{\rm RX}\P^{\rm H}_{\rm RX}\N,
\end{align}
where $\Y = [\y(1),\ldots,\y(T)]$, $\S \triangleq [\s(1),\ldots,\s(T)]$, and $\N \triangleq [\n(1),\ldots,\n(T)]$, with each $\n(t)\sim\mathcal{CN}(\mathbf{0}_{N_{\rm RF}\times1},\sigma^2\mathbf{I}_{N_{\rm RF}})$ being the $t$th slot AWGN vector. Finally, $\H_{\rm SI}\in \mathbb{C}^{N \times N}$ is the SI matrix~\cite{alexandropoulos2017joint,FD_HMIMO_2023} and $\D \in \mathbb{C}^{N_{\rm RF} \times N_{\rm RF}}$ denotes the digital SI cancellation matrix at the FD XL MIMO node.
%

\section{Proposed FD-Enabled XL MIMO ISAC}
In this section, we deploy the aforedescribed FD XL MIMO system for simultaneous multi-user data communications and monostatic sensing of multiple targets. We particularly present a joint design of the TX/RX DMAs' analog BF matrices, the TX digital BF matrix, and the digital SI cancellation matrix for DL sum-rate maximization, while ensuring a minimum requirement for the targets' 3D parameter estimation.

\subsection{Problem Formulation}
Our goal is to optimize the parameters of the FD XL MIMO system for DL sum-rate maximization under a targets' PEB constraint ensuring their spatial parameters' estimation accuracy. To this end, we formulate the optimization problem:
\begin{align}
        \mathcal{OP}&:\nonumber\underset{\substack{\W_{\rm TX},\W_{\rm RX}\\ \V,\D}}{\max} \quad \sum\limits_{u=1}^{U}\log\Big(1+\left|\widehat{\h}_{{\rm DL},u}\P_{\rm TX}\W_{\rm TX}\v_u\right|^2\sigma^{-2}\Big)\\
        &\nonumber\text{\text{s}.\text{t}.}\,
        \left\|[\W^{\rm H}_{\rm RX}\P^{\rm H}_{\rm RX}\H_{\rm SI}\P_{\rm TX}\W_{\rm TX}\V]_{(i,:)}\right\|^2\leq \gamma_{\rm SI}\, \forall i,\\
        &\nonumber\,\quad \sum\limits_{u=1}^{U}\left\|\P_{\rm TX}\W_{\rm TX}\v_u\right\|^2 \leq P_{\rm max},\,\, w^{\rm TX}_{i,n}, w^{\rm RX}_{i,n} \in \mathcal{W},\\
        &\nonumber\,\,\quad {\rm PEB}(\widehat{\boldsymbol{\zeta}}|\W_{\rm RX},\W_{\rm TX},\V)\leq\gamma_s.
\end{align}
In this formulation, the first constraint refers to the residual SI signal at the output of each RX DMA's microstrip, while the last constraint $\gamma_s$ deals with the PEB performance, which is expressed as a function of the analog TX/RX and digital TX BF matrices. To construct the estimations for the involved channels, we use vector $\widehat{\boldsymbol{\zeta}}\triangleq[(\widehat{r}_1,\widehat{\theta}_1,\widehat{\phi}_1),\ldots,(\widehat{r}_U,\widehat{\theta}_U,\widehat{\phi}_U)]^{\rm T}$ with the estimation of the targets' 3D parameters: $\widehat{\h}_{{\rm DL},u}$ $\forall u$ is composed using~\eqref{eqn:DL_chan} while the composite end-to-end channel $\widehat{\H}_{\rm R}$ via \eqref{eq:H_R}, excluding the coefficients $\beta_k$'s which are unknown. $\mathcal{OP}$ is a highly coupled and non-convex problem, making the search for an optimal solution computationally demanding. Therefore, we next pursue a sub-optimal approach focused on first determining separately the sensing- and communication-oriented system parameters, which are then combined to formulate a configuration that aligns with $\mathcal{OP}$'s ISAC criterion. In the subsequent sections, we present the sensing- and communication-oriented designs.

\subsubsection{PEB Analysis}
It is evident from \eqref{eq:received_matrix}'s inspection that, for a coherent communication block involving a sufficiently large number of transmissions, we can assume that $\frac{1}{T}\S\S^{\rm H}=\I_{U}$ and that the received signal at the output of the RX DMA $\Y\sim\mathcal{CN}(\boldsymbol{\mu},\R_n)$, with mean $\boldsymbol{\mu} \triangleq \W^{\rm H}_{\rm RX}\P^{\rm H}_{\rm RX}\H_{\rm R}\P_{\rm TX}\W_{\rm TX}\V\S$ and covariance matrix $\R_n \triangleq \sigma^2(\W_{\rm RX}^{\rm H}\P_{\rm RX}^{\rm H})(\P_{\rm RX}\W_{\rm RX})\I_{N_{\rm RF}}$. In the context of estimating $\boldsymbol{\zeta} \triangleq [(r_1,\theta_1,\phi_1),\ldots,(r_U,\theta_U,\phi_U)]^{\rm T}$ (i.e., $\widehat{\boldsymbol{\zeta}}$'s computation), each $(i,j)$-th element of its $3U\times3U$ Fisher Information Matrix (FIM) can be calculated as follows~\cite{kay1993fundamentals}:
\begin{align}
    \nonumber\text{diag}(\mathbfcal{I}) = (\mathbfcal{I}_{r_1r_1},\mathbfcal{I}_{\theta_1\theta_1},\mathbfcal{I}_{\phi_1\phi_1},\ldots,\mathbfcal{I}_{r_Ur_U},\mathbfcal{I}_{\theta_U\theta_U},\mathbfcal{I}_{\phi_U\phi_U})
\end{align}
with its diagonal elements defined as:
\begin{align}
    \nonumber[\mathbfcal{I}]_{i,j} = 2\Re\left\{\frac{\partial \boldsymbol{\mu}^{\rm H}}{\partial\zeta_i}\R_n^{-1}\frac{\partial \boldsymbol{\mu}}{\partial\zeta_j}\right\}+\text{Tr}\left(\R_n^{-1}\frac{\partial\R_n}{\partial\zeta_i}\R_n^{-1}\frac{\partial\R_n}{\partial\zeta_j}\right).
\end{align}
It is apparent that $\nabla_{\boldsymbol{\zeta}}\R_n = \mathbf{0}_{3U\times1}$ because $\R_n$ is independent of $\boldsymbol{\zeta}$; therefore, each FIM value depends solely on the mean and can be expressed as follows $\forall\zeta_i\in\boldsymbol{\zeta}$:
\begin{align}\label{eq: deriv_mean}
    \frac{\partial\boldsymbol{\mu}}{\partial\zeta_i} = \W^{\rm H}_{\rm RX}\P^{\rm H}_{\rm RX}\frac{\partial\H_{\rm R}}{\partial\zeta_i}\P_{\rm TX}\W_{\rm TX}\V\S.
\end{align}
Consequently, each diagonal element of the FIM matrix can be re-written as:
\begin{align}
    \nonumber[\mathbfcal{I}]_{i,i} = 2\Re\Bigg\{&\S^{\rm H}\V^{\rm H}\W_{\rm TX}^{\rm H}\P_{\rm TX}^{\rm H}\frac{\partial\H_{\rm R}^{\rm H}}{\partial\zeta_i}\P_{\rm RX}\W_{\rm RX}\R_n^{-1}\\&\nonumber\W^{\rm H}_{\rm RX}\P^{\rm H}_{\rm RX}\frac{\partial\H_{\rm R}}{\partial\zeta_i}\P_{\rm TX}\W_{\rm TX}\V\S\Bigg\}.
\end{align}
Putting all above together, the PEB of $\boldsymbol{\zeta}$ can be expressed as a function of the analog TX/RX and digital TX BF matrices:
\begin{align}\label{eq: PEB}
    \nonumber{\rm PEB}({\boldsymbol{\zeta}}|\W_{\rm RX},\W_{\rm TX},\V)&\triangleq\sqrt{({\rm CRB}\left({\boldsymbol{\zeta}}|\W_{\rm RX},\W_{\rm TX},\V\right)}\\\nonumber&=\sqrt{{\rm Tr}\{\mathbfcal{I}^{-1}\}}
\end{align}

\subsubsection{Cram\'{e}r-Rao Bound Optimization}
We now focus on designing the TX/RX DMAs' analog and TX DMA's digital BF matrices for minimizing $\boldsymbol{\zeta}$'s CRB. To this end, we leverage the positive semidefinite nature of the FIM and the lower bound ${\rm Tr}\{\mathbfcal{I}^{-1}\}\geq\frac{9U^2T^2}{{\rm Tr}\{\mathbfcal{I}\}}$ resulting from the previously derived PEB expression. In mathematical terms, we formulate the  following optimization problem: 
\begin{align}
        \mathcal{OP}_{\rm PEB}&:\nonumber\underset{\substack{\W_{\rm TX},\W_{\rm RX},\V}}{\max} \quad {\rm Tr}\left\{\sum_{u=1}^{U}\mathbfcal{I}_{r_ur_u}+\mathbfcal{I}_{\theta_u\theta_u}+\mathbfcal{I}_{\phi_u\phi_u}\right\}\\
        &\nonumber\quad\text{\text{s}.\text{t}.}\, w^{\rm TX}_{i,n}, w^{\rm RX}_{i,n} \in \mathcal{W}.\nonumber
\end{align}
In the subsequent analysis, we disregard the codebook constraint and define $\F_{\rm RX} \triangleq \W_{\rm RX}\R^{-1}_n\W_{\rm RX}^{\rm H}$, $\F_{\rm TX} \triangleq (\W_{\rm TX}\V)(\V^{\rm H}\W_{\rm TX}^{\rm H})$, and $\K_{\rm \zeta_i}\triangleq\P_{\rm RX}^{\rm H}\frac{\partial \H_{\rm R}}{\partial \zeta_i}\P_{\rm TX}$ $\forall\zeta_i\in\boldsymbol{\zeta}$. Using these definitions, $\mathcal{OP}_{\rm PEB}$'s objective simplifies to:
\begin{align}
        \nonumber\underset{\substack{\F_{\rm TX},\F_{\rm RX}}}{\max} \quad \sum_{i=1}^{3U}{\rm Tr}\left\{\Re\{\K_{\rm \zeta_i}\F_{\rm TX}\K_{\rm \zeta_i}^{\rm H}\F_{\rm RX}\}\right\}.
\end{align}
Due to the partially-connected DMA architecture, it can be shown that $\R_n$ takes the form of a diagonal matrix with the following distinctive composition:
\begin{align}
    \nonumber\R_n = \text{diag}\{&(\w_1^{\rm H}\circ\p_1^{\rm H})(\p_1\circ\w_1), \ldots ,\\
    &(\w_{N_{\rm RF}}^{\rm H}\circ\p_{N_{\rm RF}}^{\rm H})(\p_{N_{\rm RF}}\circ\w_{N_{\rm RF}})\}\sigma^2,
\end{align}
where $\circ$ is the Hadamard product and $\forall$$i=1,\ldots,N_{\rm RF}$:
\begin{align}
    &\nonumber \p_i \triangleq [\text{diag}(\P_{\rm RX})]_{((i-1)N_{\rm E}+1:iN_{\rm E})}\\
    &\nonumber \w_i \triangleq [\W_{\rm RX}]_{((i-1)N_{\rm E}+1:iN_{\rm E},i)},
\end{align}
Consequently, it can be easily verified that $\F_{\rm RX}$ has a block diagonal structure with each block having the following form:
\begin{align}\label{eq:block}
    \frac{\w_i\w_i^{\rm H}}{(\w_i^{\rm H}\circ\p_i^{\rm H})(\p_i\circ\w_i)\sigma^2} \,\quad \forall i=1,\ldots,N_{\rm RF}.
\end{align}

The problem $\mathcal{OP}_{\rm PEB}$ is non-convex due to the structure of $\F_{\rm RX}$ and $\F_{\rm TX}$. To tackle this inherent non-convexity in the objective function, we next employ positive semidefinite relaxation. In addition, due to $\F_{\text{RX}}$'s block-diagonal structure, $\mathcal{OP}_{\rm PEB}$ can be decoupled into the following two distinct optimization problems that can solved via alternating optimization until convergence: 
\begin{align}
        \mathcal{OP}_{\rm PEB}^{\rm RX}&:
        \nonumber\underset{\substack{\{\F_1,\ldots,\F_{N_{\rm RF}}\}}}{\max} \quad{\rm Tr}\left\{\Re\left\{\sum_{i=1}^{3U}(\K_{\rm \zeta_i}\F_{\rm TX}\K_{\rm \zeta_i}^{\rm H})\F_{\rm RX}\right\}\right\}\\
        &\nonumber\quad\,\,\text{\text{s}.\text{t}.}\, \F_{\rm RX}=\text{diag}(\F_1,\ldots,\F_{N_{\rm RF}}),\,\,\F_{\rm RX}\succeq0,\\\nonumber\\
        \mathcal{OP}_{\rm PEB}^{\rm TX}&:
        \nonumber\underset{\substack{\F_{\rm TX}}}{\max} \quad{\rm Tr}\left\{\Re\left\{\sum_{i=1}^{3U}\F_{\rm TX}(\K_{\rm \zeta_i}^{\rm H}\F_{\rm RX}\K_{\rm \zeta_i})\right\}\right\}\\
        &\nonumber\quad\text{\text{s}.\text{t}.}\, \F_{\rm TX}\succeq0.
\end{align}
Once convergence is achieved, we conduct codebook similarity checks for each block $\F_i$ within $\F_{\rm RX}$. This involves minimizing the distance function $d(\F_i,\F_j)=\|\F_i-\F_j\|_F$ for every corresponding $\F_j = \frac{\w_j\w_j^{\rm H}}{(\w_j^{\rm H}\circ\p_i^{\rm H})(\p_i\circ\w_j)\frac{\sigma^2}{T}}$ with $\w_j\in\mathcal{W}$. In doing so, we seek the corresponding codebook-constrained solution for $\mathcal{OP}_{\rm PEB}^{\rm RX}$ and construct $\W_{\rm RX}$ accordingly. Note that, due to $\F_{\rm RX}$'s block-diagonal structure, parallelization can be utilized. Then, for determining the corresponding $\W_{\rm TX}$ and $\V$, we focus on the following matrix factorization problem:
 \begin{align}
        \mathcal{OP}_{\rm FACT}&:
        \nonumber\underset{\substack{\W_{\rm TX},\V}}{\max} \|\F_{\rm TX}-\W_{\rm TX}\V\V^{\rm H}\W_{\rm TX}^{\rm H}\|_F^2\\
        &\nonumber\quad\text{\text{s}.\text{t}.}\, \sum\limits_{u=1}^{U}\|\P_{\rm TX}\W_{\rm TX}\v_u\|^2 \leq P_{\rm max}, \,\, w^{\rm TX}_{i,n}\in\mathcal{W}.
\end{align}
To solve this problem, we iteratively perform a sequence of $N_{\rm RF}$ 1D searches to pinpoint the $\W_{\rm TX}$ that minimize the objective, which is coupled with nonlinear least-squares fitting to determine the corresponding $\V$. This iterative process continues until convergence is achieved. It is worth noting that the optimization methods described can be efficiently tackled using convex optimization solvers (e.g., CVX \cite{cvx}) and optimization tools available in mathematical software.

\subsubsection{Downlink Sum-Rate Maximization}
The communication-optimizing configuration of the FD XL MIMO system focuses on designing the TX DMA's analog and digital BF matrices for the DL sum-rate maximization. In mathematical terms:
\begin{align}
        \mathcal{OP}_{\rm COM}&:\nonumber\underset{\substack{\W_{\rm TX},\V}}{\max} \,\, \sum\limits_{u=1}^{U}\log\Big(1+\left|\widehat{\h}_{{\rm DL},u}\P_{\rm TX}\W_{\rm TX}\v_u\right|^2\sigma^{-2}\Big)\\
        &\quad\nonumber\text{\text{s}.\text{t}.}\,\, \sum_{u=1}^U\left\|\P_{\rm TX}\W_{\rm TX}\v_u\right\|^2\leq P_{\rm max},\,\, w^{\rm TX}_{i,n} \in \mathcal{W}.
\end{align}
This problem can be efficiently solved via the following decoupled way. Firstly, the matrix $\W_{\rm TX}$ can be obtained via $N_{\rm RF}$ 1D searches within the available codebook $\mathcal{W}$. Then, the matrix $\V$ can designed to implement block diagonalization~\cite{FD_HMIMO_2023}, thus, maximizing the Signal-to-Noise Ration (SNR) for the UEs in the DL while minimizing inter-UE interference.

\subsection{Proposed ISAC Design}
After having identified both system configurations tailored for either communicatiosn or sensing purposes, we utilize the $\W_{\rm RX}$ solution of $\mathcal{OP}_{\rm PEB}^{\rm RX}$ for sensing and determine the TX parameter setup by identifying the TX BF configuration that aligns most efficiently with the ISAC-oriented $\mathcal{OP}$. We specifically focus on the following weighted linear combinations:
\begin{align}\label{eq: line_search}
    &\nonumber\W_{\rm TX} = \xi\W_{\rm TX,c}+(1-\xi)\W_{\rm TX,s},\quad\xi\in[0,1],\\
    &\V = \rho\V_{\rm c}+(1-\rho)\V_{\rm s},\quad\rho\in[0,1],
\end{align}
where $\W_{\rm TX,c}$ and $\V_{\rm c}$ denote the communication-oriented TX DMA analog and digital BF matrices obtained from solving $\mathcal{OP}_{\rm COM}$. Similarly, $\W_{\rm TX,s}$ and $\V_{\rm s}$ represent their counterparts resulting from $\mathcal{OP}_{\rm FACT}$'s solution, which target sensing optimization. It is noted that, in the context of the TX DMA analog BF matrix, we cannot ensure that combining two codewords linearly will result in another codeword from the codebook. To address this issue, we propose to conduct an additional 1D search to identify the best match by minimizing the norm distance, between the codebook constrained and unconstrained codewords.

The overall FD-enabled XL MIMO ISAC scheme is summarized in Algorithm~\ref{alg:the_opt}. Note that the feasibility region of $\gamma_s$ within $\mathcal{OP}$ is dependent on the combinations outlined in \eqref{eq: line_search}. Therefore, we assume the utilization of a feasible $\gamma_s$ that aligns with the potential values that the PEB may take. 

\begin{algorithm}[!t]
    \caption{FD-Enabled XL MIMO ISAC Configuration}
    \label{alg:the_opt}
    \begin{algorithmic}[1]
        \renewcommand{\algorithmicrequire}{\textbf{Input:}}
        \renewcommand{\algorithmicensure}{\textbf{Output:}}
        \REQUIRE $\P_{\rm TX}$, $\P_{\rm RX}$ and ${\H}_{\rm SI}$. 
        \ENSURE $\W_{\rm TX}$, $\W_{\rm RX,s}$, $\V$, and $\D$.
         \STATE Obtain $(\widehat{r}_k,\widehat{\theta}_k,\widehat{\varphi}_k)$ $\forall k$ as in~\cite{FD_HMIMO_2023} and construct $\widehat{\boldsymbol{\zeta}}$.
        \STATE Set $\widehat{\H}_{\rm R} = \sum\limits_{u=1}^{U} \a_{\rm RX}(\widehat{r}_u,\widehat{\theta}_u,\widehat{\varphi}_u)\a_{\rm TX}^{\rm H}(\widehat{r}_u,\widehat{\theta}_u,\widehat{\varphi}_u)$ and construct $\widehat{\h}_{{\rm DL},u}$ $\forall$$u$ using \eqref{eqn:DL_chan}.
        \STATE Initialize $\F_{\rm TX}$ and $\F_{\rm RX}$, calculate $\frac{\partial \widehat{\H}_{\rm R}}{\partial \zeta_i}$ $\forall \zeta_i\in\widehat{\boldsymbol{\zeta}}$, and construct $\K_{\zeta_i}$ accordingly.
        \STATE Alternately optimize $\mathcal{OP}_{\rm PEB}^{\rm RX}$ and $\mathcal{OP}_{\rm PEB}^{\rm TX}$ until convergence to obtain $\F_{\rm RX}$ and $\F_{\rm TX}$.
        \STATE Obtain $\W_{\rm RX}$ from $\F_{\rm RX}$ via \eqref{eq:block} as well as $\W_{\rm TX,s}$ and $\V_s$ solving $\mathcal{OP}_{\rm FACT}$.
        \STATE Solve $\mathcal{OP}_{\rm COM}$ to acquire $\W_{\rm TX,c}$ and $\V_c$.
        \STATE Search over the possible combinations of $\W_{\rm TX}$ and $\V$ in \eqref{eq: line_search} to find the one that best fits $\mathcal{OP}$.
        \STATE Set $\D = -(\W^{\rm H}_{\rm RX,s}\P_{\rm RX}^{\rm H}\H_{\rm SI}\P_{\rm TX}\W_{\rm TX})$ and design $\F$ as the $N_{\rm RF}$ right-singular vectors of $-\D$.
        \FOR{$\alpha=N_{\rm RF}-1,N_{\rm RF}-2,\ldots,L$}
            \IF{$\|[{\W}^{\rm H}_{\rm RX}\P_{\rm RX}^{\rm H}{\H}_{\rm SI}\P_{\rm TX}{\W}_{\rm TX}\F\widetilde{\V}]_{(i,:)}\|^2\leq \gamma$ $\forall i$ and ${\rm PEB}(\widehat{\boldsymbol{\zeta}}|\W_{\rm RX},\W_{\rm TX},\F\widetilde{\V})\leq\gamma_s$}
    			 \STATE Output $\W_{\rm TX}$, $\W_{\rm RX}$, $\V=\F\widetilde{\V}$, and $\D$, and exit.
    		\ENDIF
            \STATE Set $\F_{(:,N_{\rm RF}-\alpha+1:N_{\rm RF})}=0$.
        \ENDFOR
        \STATE Output that the FD node's settings do not meet the residual SI constraint or the PEB constraint.		
    \end{algorithmic}
\end{algorithm}

\begin{figure*}[!t]
  \begin{subfigure}[t]{0.35\textwidth}
  \centering
    \includegraphics[scale=0.347]{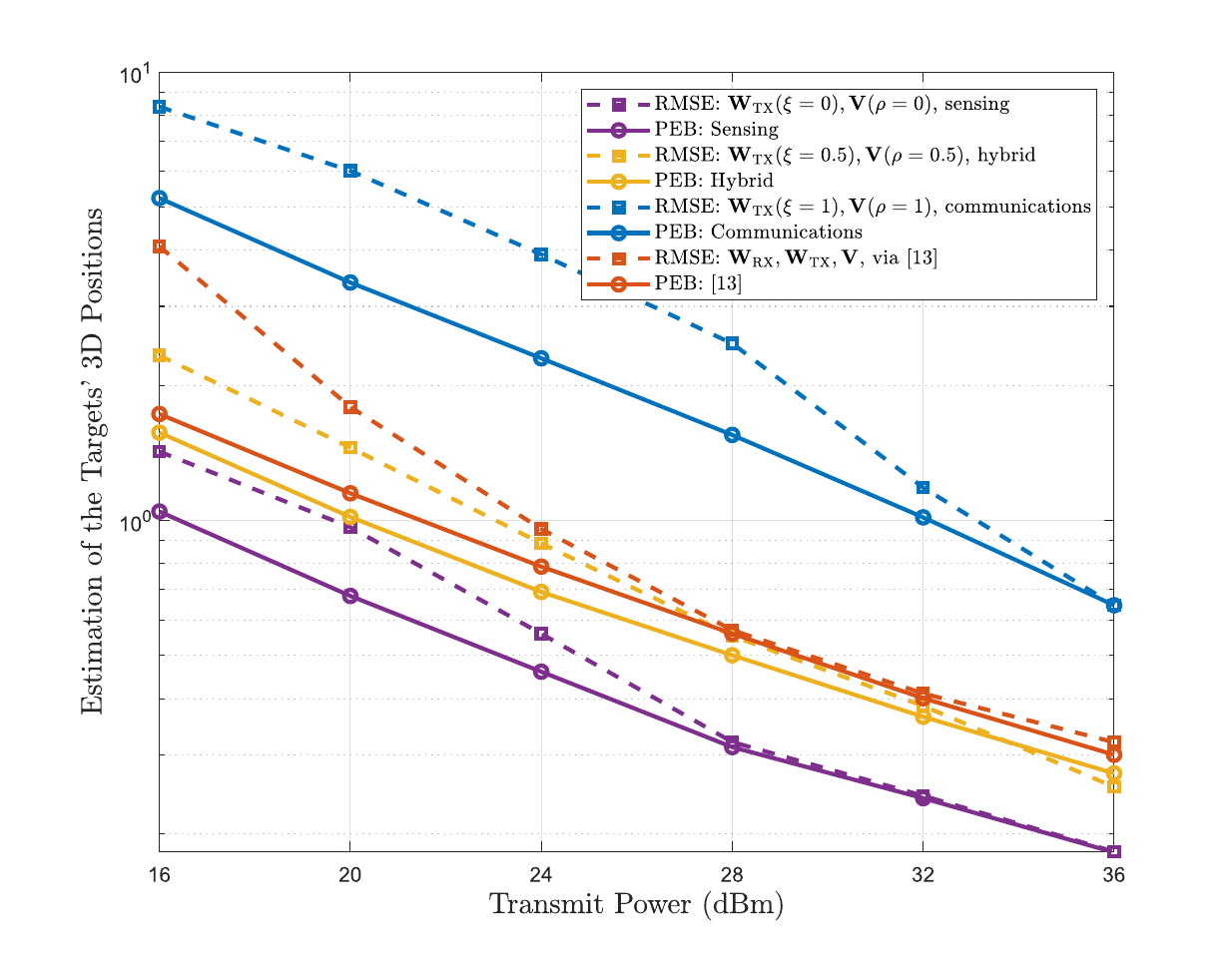}
    \caption{Sensing Performance.}
    \label{fig:RMSE}
  \end{subfigure}\hfill
  \begin{subfigure}[t]{0.35\textwidth}
  \centering
  \hspace*{-1cm}
    \includegraphics[scale=0.347]{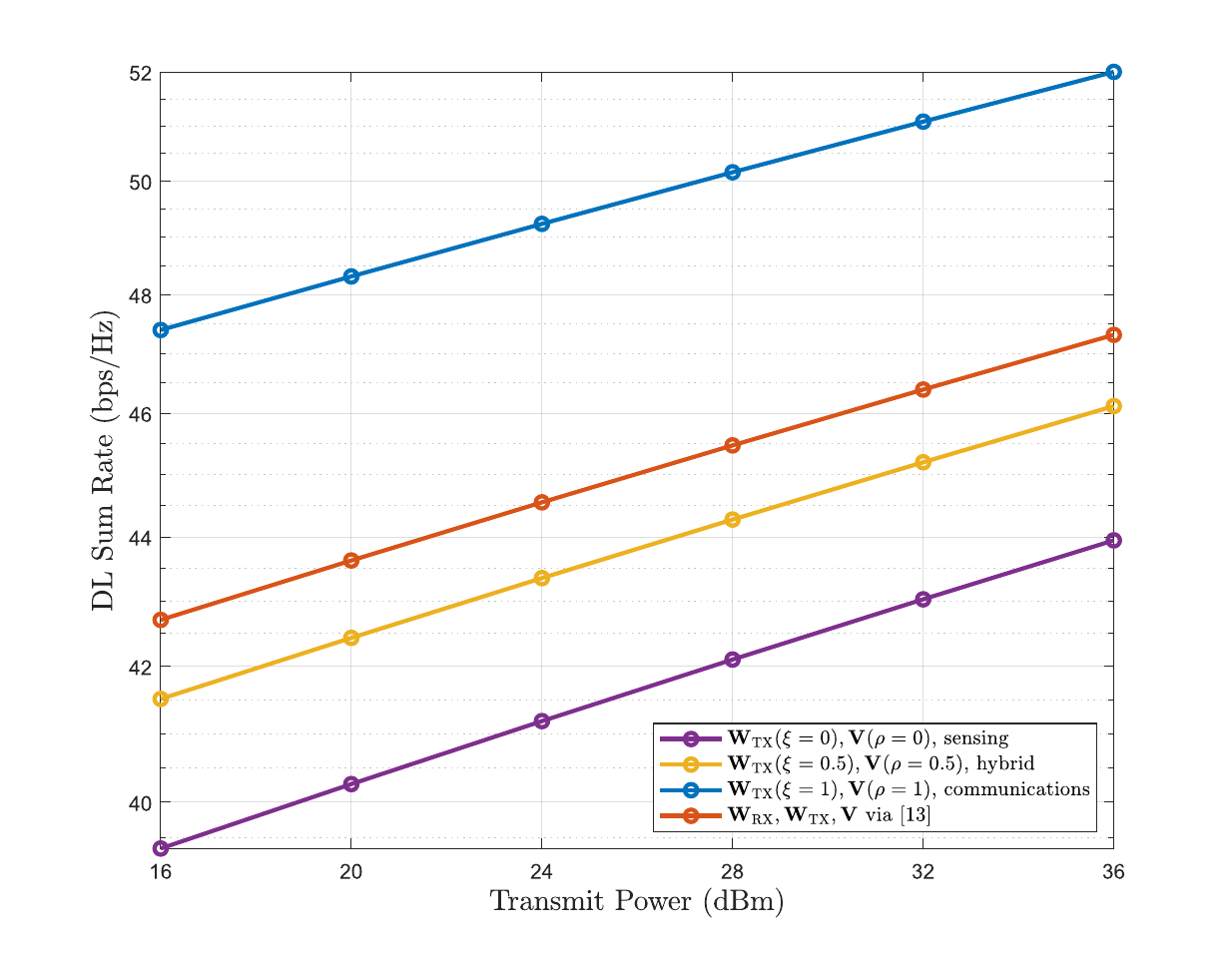}
    \caption{Communications' Performance.}
    \label{fig:DL}
  \end{subfigure}
  \caption{\small{Sensing and communications' performance for $K=3$ targets, $U=2$ of which being UEs, versus the transmit power $P_{\rm max}$ with the proposed FD-enabled XL-MIMO ISAC scheme for the case where $N_{\rm RF} = 4$ TX/RX microstrips, each with $N_{\rm E}=128$ metamaterials.}}\vspace{-0.4cm}
  \label{fig:Estimation_vs_P_T}
\end{figure*}

\section{Numerical Results}

In this section, we assess the performance of the proposed FD-enabled XL MIMO ISAC scheme for a scenario including $K=3$ targets, out of which $U=2$ are the data-requesting UEs. The FD node deploys TX/RX DMAs each with $N_{\rm RF} = 4$ microstrips, each comprising $N_{\rm E}=128$ metamaterials, while the inter- and intra-microstrip distances are set as $d_{\rm RF}=d_{\rm E}=\lambda/2$, and TX/RX DMA separation is $d_{\rm P}=0.1$ meters. The ISAC system operates at the central frequency $120$ GHz with a bandwidth of $B=150$ KHz, and the targets' parameter estimation follows the approach presented in~\cite{FD_HMIMO_2023}. Each UE location undergoes $T=200$ transmission slots, with all UEs being randomly positioned at the fixed elevation angle $30^{\circ}$, azimuth angle in $[0^{\circ},180^{\circ}]$, and range between $1$ and $15$ meters (within the Fresnel region) for $300$ Monte Carlo runs. The noise variance $\sigma^2$ in dBm was set to $-174 + 10\log_{10}(B)$ and $\beta_k$ in \eqref{eq:H_R} was randomly chosen with unit amplitude. Finally, the TX/RX DMAs' codebook $\mathcal{W}$ was designed as follows. We started with a $10$-bit DFT beam codebook $\mathcal{F}\in\{e^{j\phi}|\phi\in\left[-\pi/2,\pi/2\right]\}$ having constant amplitude and arbitrary phase values. Given any matrices $\widetilde{\W}_{\rm TX}$ and $\widetilde{\W}_{\rm RX}$ and compensating for the signal propagation inside the microstrips, the respective final TX/RX DMA weights $\W_{\rm TX}$ and $\W_{\rm RX}$ were derived as $w^{\rm TX}_{i,n} \triangleq 0.5(\jmath+\widetilde{w}^{\rm TX}_{i,n}e^{\jmath\rho_{i,n} \beta_{i}})$ and $w^{\rm RX}_{i,n} \triangleq 0.5(\jmath+\widetilde{w}^{\rm RX}_{i,n}e^{\jmath\rho_{i,n} \beta_{i}})$.

In Figs.~\ref{fig:RMSE} and \ref{fig:DL}, the performance of the proposed ISAC scheme in Algorithm~\ref{alg:the_opt} is illustrated versus $P_{\rm max}$ in dBm. In particular, Fig.~\ref{fig:RMSE} depicts the Root Mean Square Error (RMSE) for the estimation of the targets' 3D positions and the respective PEBs, while Fig. \ref{fig:DL} shows the achievable DL sum-rate performance. Multiple configurations of Algorithm~\ref{alg:the_opt} were considered and compared to \cite{FD_HMIMO_2023}'s ISAC scheme. As observed, the targets' sensing performance improves as SNR levels increase closely approaching the PEB. This fact witnesses our scheme's capability to effectively cancel out SI at the FD node. It is also shown that, even for the minimum number $N_{\rm RF}=4$ of RF chains used to localize $K=3$ targets, a purely communication-oriented configuration yields adequate localization, thanks to the sensing-oriented analog RX BF. Conversely, in a pure sensing-focused design, we can achieve a comparable DL sum rate. Finally, the hybrid design in both figures underscores our scheme's versatility across diverse scenarios. This stands in sharp contrast to the optimization framework outlined in \cite{FD_HMIMO_2023}, which offers a singular solution to a considerably more intricate problem, thus showcasing that it only provides a single configuration setting out of the many possible alternatives. The possible configurations are defined by the sensing and communication boundaries corresponding to $\xi=\rho=0$ and $\xi=\rho=1$, respectively, providing insights in the scheme's expected sensing and communication trade-off.

\section{Conclusion}
In this paper, we presented an FD-enabled XL MIMO system for simultaneous multi-user communications and multi-target sensing in the near-field regime at THz frequencies. Considering DMAs at the FD transceiver, we derived the PEB of the targets' 3D spatial parameters' estimation and proposed a suboptimal, yet computationally efficient, framework that identifies the combination between sensing- and communication-optimized system parameters that best fits a desirable ISAC criterion. The presented numerical results showcased the validity of the conducted PEB analysis as well as the combined superior PEB and DL rate performance of the proposed ISAC scheme over a state-of-the-art approach.

\bibliographystyle{IEEEtran}
\bibliography{ms}
\end{document}